\documentclass[11pt]{article}
\usepackage{latexsym}  % for Symbol fonts
\usepackage{amssymb}
\usepackage{graphicx}
\usepackage{amsmath}
\begin{document}
\hspace*{11cm} {OU-HET-652/2010}

\begin{center}
{\Large\bf Yukawaon model and unified description of}\\
{\Large\bf quark and lepton mass matrices}\footnote{
To appear in the Proceedings of Lepton-Photon 2009.}

\vspace{5mm}
{\bf Yoshio Koide}

{\it Department of Physics, Osaka University,  
Toyonaka, Osaka 560-0043, Japan} \\
{\it E-mail address: koide@het.phys.sci.osaka-u.ac.jp}

%\vspace{2mm}
\date{\today}
\end{center}

\begin{abstract}
In the so-called yukawaon model, where effective 
Yukawa coupling constants $Y_f^{eff}$ ($f=e,\nu,u,d$) are 
given by vacuum expectation values of gauge singlet 
scalars (yukawaons) $Y_f$ with $3\times 3$ flavor components, 
it is tried to give a unified description of quark and lepton
mass matrices. 
Especially, without assuming any discrete symmetry in the 
lepton sector, nearly tribimaximal mixing is derived 
by assumed a simple up-quark mass matrix form.
\end{abstract}

\section{What is a yukawaon model?}

First, let us give a short review of the so-called {\em yukawaon} model: 
We regard Yukawa coupling constants $Y_f$ as effective
coupling constants $Y_f^{eff}$ in an effective theory, 
and we consider that 
$Y_f^{eff}$ originate in vacuum expectation values (VEVs)
of new gauge singlet scalars $Y_f$, i.e.
$$
Y_f^{eff} =\frac{y_f}{\Lambda} \langle Y_f\rangle ,
\eqno(1)
$$
where $\Lambda$ is a scale of an effective theory which is 
valid at $\mu \leq \Lambda$, and we assume 
$\langle Y_f\rangle \sim \Lambda$.
We refer the fields $Y_f$ as 
{\em yukawaons} \cite{yukawaon} hereafter.  
Note that the effective coupling constants $Y_f^{eff}$ evolve
as those in the standard SUSY model below the scale $\Lambda$, 
since a flavor symmetry is completely broken at a high energy 
scale $\mu \sim \Lambda$.

In the present work, we assume an O(3) flavor symmetry.
In order to distinguish each $Y_f$ from others, 
we assume a U(1)$_X$ symmetry (i.e. {\em sector charge}). 
(The SU(2)$_L$ doublet fields $q$, $\ell$, $H_u$ and $H_d$
are assigned to sector charges $Q_X=0$.) 
Then, we obtain VEV relations as follows: 
(i) We give an O(3) and U(1)$_X$ invariant superpotential for 
yukawaons $Y_f$. 
(ii) We solve SUSY vacuum conditions $\partial W/\partial Y_f=0$.
(iii) Then, we obtain VEV relations among $Y_f$.      
 
For example, in the seesaw-type neutrino mass matrix,  
$M_\nu \propto \langle Y_\nu\rangle \langle Y_R\rangle^{-1} 
\langle Y_\nu\rangle^T$,  we obtain  \cite{Koide-O3-PLB08}
$$
\langle Y_R\rangle \propto  \langle Y_e\rangle 
\langle \Phi_u\rangle + \langle\Phi_u\rangle \langle Y_e\rangle 
\eqno(2)
$$ 
together with $\langle Y_\nu\rangle \propto \langle Y_e\rangle$
and $\langle Y_u\rangle \propto \langle\Phi_u\rangle 
\langle\Phi_u\rangle$, i.e. 
a neutrino mass matrix is given by
$$
\langle M_\nu \rangle_e \propto \langle Y_e\rangle_e 
\left\{ \langle Y_e\rangle_e \langle \Phi_u\rangle_e 
+ \langle\Phi_u\rangle_e \langle Y_e\rangle_e 
\right\}^{-1} \langle Y_e\rangle_e ,
\eqno(3)
$$
where $\langle \Phi_u\rangle_u \propto {\rm diag}
(\sqrt{m_u}, \sqrt{m_c} , \sqrt{m_t})$,
and $\langle A \rangle_f$ denotes a form of a VEV matrix
$\langle A \rangle$ in the diagonal basis of $\langle Y_f\rangle$
(we refer it as $f$ basis).
We can obtain a form $\langle \Phi_u\rangle_d = V(\delta)^T 
\langle \Phi_u\rangle_u V(\delta)$ from the definition 
of the CKM matrix $V(\delta)$, but we do not know an explicit 
form of $\langle \Phi_u\rangle_e$.
Therefore, in a previous work \cite{Koide-O3-PLB08}, 
the author put an ansatz, 
$\langle \Phi_u\rangle_e = V(\pi)^T  \langle \Phi_u\rangle_u
V(\pi)$ by supposing $\langle \Phi_u\rangle_e \simeq 
\langle \Phi_u\rangle_d$, and he obtained excellent predictions
of the neutrino oscillation parameters  
%$\sin^2 2\theta_{atm}=0.995$, $|U_{13}|=0.001$ and 
%$\tan^2\theta_{solar}=0.553$, 
without assuming any discrete symmetry.
However, there is no theoretical ground for the ansatz
for the form  $\langle \Phi_u\rangle_e$.

The purpose of the present work is to investigate a 
quark mass matrix model in order to predict neutrino
mixing parameters on the basis of a yukawaon model (2), 
without such the ad hoc ansatz,  because if we give a quark 
mass matrix model where mass matrices $(M_u, M_d)$ are given 
on the $e$ basis, then,  we can obtain the form 
$\langle \Phi_u\rangle_e$ by using a transformation 
$\langle \Phi_u\rangle_e = U_u \langle \Phi_u\rangle_u U_u^T$, 
where $U_u$ is defined by $U_u^T M_u U_u= M_u^{diag}$.   

\section{Yukawaons in the quark sector}

We assume a superpotential in the quark sector \cite{Mnu_PLB09}:
$$
W_q = \mu_u [Y_u \Theta_u] +
\lambda_u [\Phi_u \Phi_u \Theta_u] 
+ \mu_u^X [\Phi_u\Theta_u^X] +\mu_d^X [Y_d \Theta_d^X]
+ \sum_{q=u,d} \frac{\xi_q}{\Lambda} [\Phi_e 
(\Phi_{X} + a_q E ) \Phi_e \Theta_q^X] .
\eqno(4)
$$
Here and hereafter, for convenience, we denotes Tr[...] as 
[....] simply.
From SUSY vacuum conditions $\partial W/\partial \Theta_u =0$, 
$\partial W/\partial \Theta_u^X =0$ and 
$\partial W/\partial \Theta_d^X =0$, we obtain 
$\langle Y_u\rangle \propto \langle\Phi_u\rangle 
\langle\Phi_u\rangle$, 
$$
M_u^{1/2}\propto 
\langle \Phi_u \rangle_e \propto \langle \Phi_e \rangle_e 
\left(
\langle \Phi_X \rangle_e + a_u \langle E \rangle_e
\right) \langle \Phi_e \rangle _e ,
\eqno(5)
$$ 
$$
M_d \propto 
\langle Y_d \rangle_e \propto \langle \Phi_e \rangle_e 
\left(
\langle \Phi_X \rangle_e + a_d \langle E \rangle_e
\right) \langle \Phi_e \rangle _e ,
\eqno(6)
$$
respectively.  
Here, $\langle \Phi_X \rangle_e$ and 
$\langle E \rangle_e$ are given by
$$
\langle \Phi_X \rangle_e \propto 
X \equiv \frac{1}{3} \left(
\begin{array}{ccc}
1 & 1 & 1 \\
1 & 1 & 1 \\
1 & 1 & 1 
\end{array} \right), \ \ \ \ \ 
\langle E \rangle_e \propto 
{\bf 1} \equiv  \left(
\begin{array}{ccc}
1 & 0 & 0 \\
0 & 1 & 0 \\
0 & 0 & 1 
\end{array} \right) .
\eqno(7)
$$
(Note that the VEV form $\langle \Phi_X \rangle_e$ 
breaks the O(3) flavor symmetry into S$_3$.)
Therefore, we obtain quark mass matrices
$$
M_u^{1/2} \propto  M_e^{1/2} \left( {X} + 
a_u   {\bf 1} \right) M_e^{1/2} , \ \ \ 
M_d \propto  M_e^{1/2} \left( {X} + 
a_d e^{i\alpha_d}  {\bf 1} \right) M_e^{1/2} ,
\eqno(8)
$$ 
on the $e$ basis.
Note that we have assumed that the O(3) relations are valid
only on the $e$ and $u$ bases, so that $\langle Y_e\rangle$ 
and $\langle Y_u\rangle$ must be real. 

A case $a_u \simeq -0.56$ can give a reasonable up-quark
 mass ratios $\sqrt{{m_{u1}}/{m_{u2}}}=0.043$ and   
$\sqrt{{m_{u2}}/{m_{u3}}}=0.057$, which are in favor of
the observed values \cite{q-mass}
$\sqrt{{m_{u}}/{m_{c}}}=0.045^{+0.013}_{-0.010}$, and
$\sqrt{{m_{c}}/{m_{t}}}=0.060\pm 0.005$ at $\mu=M_Z$.  

\section{Yukawaons in the neutrino}

However, the up-quark mass matrix (5) failed to give reasonable 
neutrino oscillation parameter values although it can 
give reasonable up-quark mass ratios.  
Therefore, we will slightly modify the model (2) 
in the neutrino sector.

Note that the sign of the eigenvalues of $M_u^{1/2}$
given by Eq.(8) is $(+, -,+)$ for the case $a_u \simeq -0.56$. 
If we assume that the eigenvalues of $\langle \Phi_u\rangle_u$ 
must be positive, so that $\langle \Phi_u\rangle_u$ in Eq.(2) 
is replaced as $\langle \Phi_u\rangle_u \rightarrow 
\langle \Phi_u\rangle_u \cdot {\rm diag}(+1,-1,+1)$,
then, we can obtain successful results except for 
$\tan^2 \theta_{solar}$, i.e. predictions 
$\sin^2 2\theta_{atm}=0.984$ and $|U_{13}|=0.0128$ and
an unfavorable prediction $\tan^2 \theta_{solar}=0.7033$.

When we introduce a new field $P_u$ with a VEV
$\langle P_u \rangle_u \propto {\rm diag}(+1,-1,+1)$,
we must consider an existence of 
$P_u Y_e \Phi_u+\Phi_u Y_e P_u$ in addition to 
$Y_e P_u \Phi_u+\Phi_u P_u Y_e$, 
because they have the same U(1)$_X$ charges.
Therefore, we modify Eq.(2) into
$$
W_R = \mu_R [Y_R \Theta_R]
 + \frac{\lambda_R}{\Lambda}
\left\{ [(Y_e P_u \Phi_u+\Phi_u P_u Y_e) \Theta_R] 
+\xi [(P_u Y_e \Phi_u +\Phi_u Y_e P_u) \Theta_R] \right\},
\eqno(9)
$$
which leads to VEV relation 
$Y_R \propto Y_e P_u \Phi_u + \Phi_u P_u Y_e 
+\xi (P_u Y_e \Phi_u + \Phi_u Y_e P_u)$.
The results at $a_u \simeq -0.56$ are excellently in favor
of the observed neutrino oscillation parameters
by taking a small value of $|\xi|$ (see Table 1):                           

Also, we can calculate the down-quark sector.
We have two parameters $(a_d, \alpha_d)$ in the 
down-quark sector given in Eq.(8). 
(See Table 2 in Ref.\cite{Mnu_PLB09}).
The results are roughly reasonable, although $|V_{i3}|$
and $|V_{3i}|$ are somewhat larger than the observed
values.  Those discrepancies will be improved in
 future version of the model.

%\begin{wraptable}{r}{0.75\textwidth}
\begin{table}
\begin{tabular}{|ccc|} \hline
Sector & Parameters & Predictions \\ \hline
      &       &  $\sin^2 \theta_{atm}$ \ \ \  $\tan^2 \theta_{solar}$
\ \ \  $|U_{13}|$ \ \\
$M_\nu$ & $\xi=+0.0005$ & $0.982$ \ \ \ \ \ 
$0.449$ \ \ \ \ \ \ \ \ $0.012$ \\ 
  & $\xi=-0.0012$ & $0.990$ \ \ \ \ \ \ 
$0.441$ \ \ \ \ \ \ \ \ $0.014$ \\ 
\cline{2-3}
$M_u^{1/2}$ & $a_u=-0.56$ & $\sqrt{\frac{m_u}{m_c}}=0.0425$ \ \ 
$\sqrt{\frac{m_c}{m_t}}=0.0570$ \\ \cline{2-3}
     & two parameters & 
 5 observables: fitted excellently \\
\hline
$M_d$ & $a_d e^{i\alpha_d}$ &  $\sqrt{\frac{m_d}{m_s}}$, \ 
$\sqrt{\frac{m_s}{m_b}}$, \ $|V_{us}|$, \ $|V_{cb}|$, \ $|V_{ub}|$, \ 
$|V_{td}|$ \\ \cline{2-3}
    & two parameters   & 6 observables: not always excellent \\ \hline
\end{tabular}
\caption{Summary of the present model.}
\label{tab:1}
%\end{wraptable}
\end{table}

\section{Summary}

In conclusion, for the purpose of deriving the observed 
nearly tribimaximal neutrino mixing, a possible yukawaon 
model in the quark sector is investigated.   
Five observable quantities (2 up-quark mass ratios and 
3 neutrino mixing parameters) are excellently fitted by 
two parameters. 
Also, the CKM mixing parameters and down-quark mass ratios 
are given under additional 2 parameters. 
The results are summarized in Table 1.

It is worthwhile to notice that the observed tribimaximal 
mixing in the neutrino sector is substantially obtained 
from the up-quark mass matrix structure (8).  
Although the model for down-quark sector still need an 
improvement, the present approach will provide a new 
view to a unified description of the masses and mixings.

% ****************************************************************************
% BIBLIOGRAPHY AREA
% ****************************************************************************

\begin{footnotesize}
% IF YOU DO NOT USE BIBTEX, USE THE FOLLOWING SAMPLE SCHEME FOR THE REFERENCES
% ----------------------------------------------------------------------------

% ----------------------------------------------------------------------------

% IF YOU USE BIBTEX,
% - DELETE THE TEXT BETWEEN THE TWO ABOVE DASHED LINES
% - UNCOMMENT THE NEXT TWO LINES AND REPLACE 'Name_Of_Your_BibFile'

%\bibliographystyle{unsrt}
%\bibliography{Name_Of_Your_BibFile}

\begin{thebibliography}{99}
%------- replace following references ;-)
\bibitem{yukawaon} Y.~Koide, Phys.~Rev. {\bf D78} 093006 (2008); 
Phys.~Rev. {\bf D79} 033009 (2009).
%
%\bibitem{K-mass-Ma} E.~Ma, Phys.~Lett.~{\bf B649}, 287 (2007).
%
\bibitem{Koide-O3-PLB08} Y.~Koide, Phys.~Lett. {\bf B665} 227 (2008).
%
%
%\bibitem{e-Yukawaon-PRD09} Y.~Koide, Phys.~Rev. {\bf D79}, 033009 (2009).
%
%\bibitem{e-Yuyakaon0902} Y.~Koide, arXiv:0902.4501.
%
%\bibitem{K-mass90} Y.~Koide, Mod.~Phys.~Lett. {\bf A5}, 2319 (1990).
%
%\bibitem{CKM_SD}
%L.~-L.~Chau and W.~-Y.~Keung, Phys.~Rev.~Lett. {\bf 53}, 1802 (1984);
%H.~Fritzsch, Phys.~Rev. {\bf D32} (1985) 3058.
%
%\bibitem{evol}  N.~Li and B.Q. Ma, Phys.~Rev. {\bf D73}, 013009 (2006);
%Z.Z.~Xing and H.~Zhang, Phys.~Lett. {\bf B635}, 107 (2006).
%
%\bibitem{PDG08} C.~Amsler, {\it et al}., Particle Data Group, 
%Phys.~Lett. {\bf B667} (2008) 1.
%
\bibitem{Mnu_PLB09} Y.~Koide, Phys.~Lett. {\bf B680} 76 (2009).
%\bibitem{tribi} 
%P.~F.~Harrison, D.~H.~Perkins and W.~G.~Scott,
% Phys.~Lett. {\bf B458} (1999)  79;
% Phys.~Lett. {\bf B530} (2002) 167;
%Z.-z.~Xing, Phys.~Lett. {\bf B533} (2002) 85;
%P.~F.~Harrison and W.~G.~Scott,  Phys.~Lett. {\bf B535} (2002) 163;
%Phys.~Lett. {\bf B557} (2003)76;
%E.~Ma, Phys.~Rev.~Lett. {\bf 90} (2003) 221802;
%C.~I.~Low and R.~R.~Volkas, Phys.~Rev. {\bf D68} (2003) 033007.
%
%\bibitem{democ-seesaw} Y.~Koide and H.~Fusaoka, Z.~Phys. {\bf C71}
% (1996) 459; Prog.~Theor.~Phys. {\bf 97}, (1997) 459.
%
%\bibitem{e-Yukawaon09} Y.~Koide, arXiv:0906.3370 [hep-ph].
%
%\bibitem{noYnu} Y.~Koide, arXiv:0812.3203 [hep-ph], a talk at  
%``Particle Physics, Astrophysics and Quantum Field Theory: 75 Years 
%since Solvay" (PAQFT08), 27-29 Nov. 2008, Nanyang Executive Centre, Singapore,
%To appear in the Conference Proceedings (Intl.J.Mod.Phys.A).
%\bibitem{Koide-U3-PLB08} Y.~Koide, Phys.~Lett. {\bf B662}, 43 (2008). 
%\bibitem{Sumino09} Y.~Sumino,  Phys.~Lett. {\bf B671}, 477 (2009). 
%
%\bibitem{polemass} 
%H.~Arason {\it et al}., Phys.~Rev. {\bf D46}, 3945 (1992).
%
\bibitem{q-mass} Z.-z.~Xing, H.~Zhang and S.~Zhou, 
Phys.~Rev. {\bf D77} (2008) 113016.
Also, see H.~Fusaoka and Y.~Koide, Phys. Rev. 
{\bf D57} (1998) 3986.
%
%\bibitem{mu-mc} Y.~Koide, Mod.~Phys.~Lett. {\bf A8}, 2071 (1993).
%
%\bibitem{MINOS} D.~G.~Michael {\it et al.}, MINOS collaboration,
%Phys.~Rev.~Lett. {\bf 97} (2006) 191801;
%J.~Hosaka, {\it et al.}, Super-Kamiokande collaboration, Phys.~Rev. 
%{\bf D74} (2006) 032002.
%
%\bibitem{SNO08} B.~Aharmim, {\it et al.}, SNO collaboration,
%Phys.~Rev.~Lett. {\bf 101} (2008) 111301.
%Also, see S.~Abe, {\it et al.}, KamLAND collaboration,
%Phys.~Rev.~Lett. {\bf 100} (2008) 221803.
%
\end{thebibliography}
% example of Name_Of_Your_BibFile.bib
% @Article{Turcato:2006ch,
%      author    = "Turcato, M.",
%  collaboration = "ZEUS and H1",
%      title     = "Lepton flavour violation and charmonium physics at HERA",
%      journal   = "Nucl. Phys. Proc. Suppl.",
%      volume    = "162",
%      year      = "2006", 
%      pages     = "283-287",
%      SLACcitation  = "%%CITATION = NUPHZ,162,283;%%"
% }
% 
% @Unpublished{Gogitidze:2007du,
%      author    = "Gogitidze, N.",
%  collaboration = "H1", 
%      title     = "Prompt photons and particle momentum distributions at
%                   HERA", 
%      year      = "2007",
%      note    = "hep-ex/0701033",
%      SLACcitation  = "%%CITATION = HEP-EX 0701033;%%"
% }

\end{footnotesize}

% ****************************************************************************
% END OF BIBLIOGRAPHY AREA
% ****************************************************************************

\end{document}